\documentclass[aps,twocolumn,prd,groupedaddress,nofootinbib,amssymb,eqsecnum,epsfig]{revtex4}
\usepackage{graphicx}
\usepackage{bm}
\usepackage{amsmath}
\usepackage{color}
\usepackage{amsfonts}

\begin{document}
\newcommand{\newc}{\newcommand}

\newc{\be}{\begin{equation}}
\newc{\ee}{\end{equation}}
\newc{\ba}{\begin{eqnarray}}
\newc{\ea}{\end{eqnarray}}
\newc{\bea}{\begin{eqnarray*}}
\newc{\eea}{\end{eqnarray*}}
\newc{\D}{\partial}
\newc{\ie}{{\it i.e.} }
\newc{\eg}{{\it e.g.} }
\newc{\etc}{{\it etc.} }
\newc{\etal}{{\it et al.}}
\newcommand{\nn}{\nonumber}
\newc{\ra}{\rightarrow}
\newc{\lra}{\leftrightarrow}
\newc{\lsim}{\buildrel{<}\over{\sim}}
\newc{\gsim}{\buildrel{>}\over{\sim}}
\def\mpl{M_{\rm pl}}
\def\d{\mathrm{d}}

\newcommand{\jose}{\textcolor{blue}}
\newcommand{\RN}{\textcolor{red}}

\title{Instabilities in Horndeski Yang-Mills inflation}

\author{Jose Beltr\'an Jim\'enez$^{1}$,
Lavinia Heisenberg$^{2}$, 
Ryotaro Kase$^{3}$, Ryo Namba$^{4}$, 
Shinji Tsujikawa$^{3}$}

\affiliation{
$^1$Aix Marseille Univ, Universit\'e de Toulon, CNRS, CPT, Marseille, France\\
$^2$Institute for Theoretical Studies, ETH Zurich, 
Clausiusstrasse 47, 8092 Zurich, Switzerland\\
$^3$Department of Physics, Faculty of Science, Tokyo University of Science, 
1-3, Kagurazaka,
Shinjuku-ku, Tokyo 162-8601, Japan\\
$^4$Department of Physics, McGill University, Montreal, QC, H3A 2T8, Canada}

\date{\today}

\begin{abstract}

A non-abelian $SU(2)$ gauge field with a non-minimal 
Horndeski coupling to gravity gives rise to a de Sitter 
solution followed by a graceful exit to a radiation-dominated 
epoch.  In this Horndeski Yang-Mills (HYM) theory we derive the second-order 
action for tensor perturbations on the 
homogeneous and isotropic quasi de Sitter background.  
We find that the presence of the Horndeski non-minimal coupling 
to the gauge field inevitably introduces ghost instabilities in the tensor sector 
during inflation.
Moreover, we also find Laplacian instabilities for the tensor perturbations deep 
inside the Hubble radius 
during inflation. Thus, we conclude that the HYM theory 
does not provide a consistent inflationary framework due to 
the presence of ghosts and Laplacian instabilities.

\end{abstract}


\maketitle

%%%%%%%%%%%%%%%%%%%%%%%%%%%%%
\section{Introduction}
%%%%%%%%%%%%%%%%%%%%%%%%%%%%%

The inflationary paradigm naturally addresses
flatness, horizon and monopole problems of 
the hot Big Bang cosmology \cite{infpapers}.  
The simplest model of inflation is based on a canonical 
scalar field slowly rolling in a nearly flat potential and producing a quasi-de Sitter phase in the early universe \cite{chaotic,natural}. A remarkable feature of this scenario is that, in addition to solving the aforementioned problems, it provides a natural mechanism for the generation of primordial density perturbations from 
quantum fluctuations of the scalar field, which are stretched out
over super-Hubble scales during inflation and eventually become large-scale density perturbations responsible for the structures observed in the universe. 
The CMB temperature anisotropies observed by 
the WMAP \cite{WMAP} and Planck \cite{Planckinf} satellites are compatible 
with the theoretical predictions of  
single scalar-field inflation, albeit with tension with some simplest realizations 
like chaotic inflation \cite{chaotic}.

Cosmological observations are consistent with an isotropic universe on large scales and, thus, models of inflation preserving isotropy seem to be observationally favored. This makes scalar fields ideal candidates for inflationary models, but other realizations of isotropic inflation are also possible, for instance those based on vector fields. If the vector field only has a temporal component, isotropy is trivially realized and accelerated isotropic solutions are possible. In the framework of traditional vector-tensor theories, accelerated solutions can be obtained easily \cite{Koivisto:2008xf,Jimenez:2009ai}. These theories generically contain instabilities, although there is some room for healthy models \cite{Jimenez:2008sq,ArmendarizPicon:2009ai,EspositoFarese:2009aj,Fleury:2014qfa}. In the class of theories generalizing the Proca action \cite{Jimenez:2014rna,Tasinato, Heisenberg,Fleury:2014qfa,Hull,Allys,Jimenez16,HKT,Kimura} with non-gauge invariant derivative self-interactions but keeping three propagating polarizations for the vector field, it is also possible to obtain de Sitter solutions from a non-trivial temporal component \cite{Tasinato, Jimenez:2016opp,GPcosmo}. Since the temporal component in these theories is non-dynamical, additional ingredients are necessary to end inflation and, thus, these models might be better suited for dark energy.

On the other hand, although one would naively think that space-like 
vector fields are incompatible with an isotropic cosmological 
background, it is in fact possible to maintain the rotational 
invariance by introducing three orthogonal vector fields 
aligned with the three spatial directions, as considered for instance in Refs.~\cite{Bento:1992wy,ArmendarizPicon:2004pm}. More precisely, this amounts to trading the broken spatial rotations due to the presence of the vectors for an internal rotational symmetry so that a diagonal global $SO(3)$ group remains unbroken. This configuration was proposed in Ref.~\cite{vectorinf} for an inflationary model with three massive vector fields $A_\mu^a$ non-minimally coupled to the Ricci scalar as $RA^a_{\mu}A^{a \mu}$.
Later, it was shown that cosmological solutions 
with accelerated expansions supported by spacelike vector fields 
coupled to the Ricci scalar are prone to either ghost or 
Laplacian instabilities \cite{Peloso}. 
In the presence of a scalar field $\phi$ coupled to the vector 
field $A_{\mu}$ of the form 
$f(\phi)^2 F_{\mu \nu}F^{\mu \nu}$ (where 
$F_{\mu \nu}=\partial_{\mu}A_{\nu}-\partial_{\nu}A_{\mu}$), 
it is possible to realize a stable inflationary solution 
with a small anisotropic hair \cite{aniinf}. 

If the vector fields are provided with a non-abelian $SU(2)$ gauge symmetry, it is possible to realize inflation followed by a reheating stage, 
dubbed {\it gauge-flation}, without the aforementioned instabilities \cite{gaugeinf}.
The rotational symmetry in 3-dimensional space can be retained 
by introducing three gauge fields $A^a_{\mu}$ with an internal $SU(2)$ 
gauge transformations \cite{Sodareview}, 
where the indices $a,b,...$ and $\mu, \nu,...$  
denote the gauge algebra and the space-time, respectively. 
Besides the standard Yang-Mills (YM) term 
$-F^{a}_{\mu \nu}F^{a \mu \nu}/4$, this model contains 
a higher-order term of the form 
$\kappa (\epsilon^{\mu \nu \lambda \sigma}F^a_{\mu \nu}F^a_{\lambda \sigma})^2$, 
where $\epsilon^{\mu \nu \lambda \sigma}$ is the totally 
anti-symmetric Levi-Civita tensor and the field strength $F$ 
is defined by 
\be
F^a_{\mu\nu} = \partial_\mu A^a_\nu -\partial_\nu A^a_\mu 
- g \epsilon^{abc}A^b_\mu A^c_\nu\,,
\label{Fmunu}
\ee
with $g$ being the gauge coupling and $\epsilon^{abc}$ the structure constants of the $SU(2)$ group.
It will be important for later to notice that these interactions do not introduce additional derivatives per field so that the equations of motion remain of second order, which means that no 
Ostrogradski instabilities arise in standard gauge-flation.
One nevertheless needs to impose $\kappa>0$ to avoid the appearance 
of ghosts on de Sitter backgrounds. 
Notice that the gauge symmetry prevents the propagation of longitudinal modes and, thus, these models will be oblivious to the potential instabilities discussed in Refs.~\cite{Peloso} for some specific models involving massive vector fields.

Gauge-flation shares a common property with Chromo-natural 
inflation \cite{Chromo} in which a pseudo-scalar field $\phi$
with a given potential is coupled to the non-abelian gauge field \cite{Adshead}. 
After the pseudo-scalar is integrated out around the potential 
minimum, one recovers gauge-flation plus small corrections.
This equivalence is sufficiently accurate at the level of both 
background and perturbations \cite{Namba}. 
Although these two models are theoretically consistent, 
the scalar spectral index $n_s$ and the tensor-to-scalar ratio $r$ 
are incompatible with those constrained from the Planck 
CMB data \cite{Adshead:2013nka,Namba}. However, one can extend the models to bring them back into accordance with CMB observations, which was done for instance in 
Refs.~\cite{Obata:2016tmo,Dimastrogiovanni:2016fuu,Adshead:2016omu} as extensions 
of the original Chromo-natural model.
Within the framework of gauge-flation, it remains to be explored whether more general Lagrangians can be compatible with observations, e.g.,~its massive variant \cite{Nieto:2016gnp}.

The gauge models discussed above only include minimal couplings of the gauge fields, but non-minimal couplings are also possible \cite{Balakin,Davydov}. However, couplings to the curvature that preserve the gauge symmetry are very delicate since one quickly runs into problems with Ostrogradski instabilities. This happens  because a coupling of the schematic form $RFF$ will typically lead to terms in the equations containing second derivatives of $F$. These terms will increase the order of the field equations and, thus,  they generically will give rise to the Ostrogradski instability. Only if the couplings are chosen with care, this instability can be avoided. For the abelian gauge field, Horndeski showed in 1976 \cite{Horndeski76} that there is a unique non-minimal coupling leading to second-order field equations\footnote{This result also follows from the fact that no Galileon-like derivative self-interactions are possible for gauge fields in four dimensions \cite{Mukoh}.}, which is of the form $L^{\mu \nu \alpha \beta}F_{\mu \nu}F_{\alpha \beta}$, with 
$L^{\mu \nu \alpha \beta}$ denoting the double dual Riemann tensor defined by 
\be
L^{\mu \nu \alpha \beta}
=-\frac{1}{2} 
\epsilon^{\mu \nu \rho \sigma}
\epsilon^{\alpha \beta \gamma \delta} 
R_{\rho \sigma \gamma \delta}\,,
\label{double}
\ee
where $R_{\rho \sigma \gamma \delta}$ is the Riemann tensor. 
For the non-abelian case, the generalization of the above non-minimal 
coupling to $L^{\mu \nu \alpha \beta}F^a_{\mu \nu}F^a_{\alpha \beta}$ 
is again the unique  non-minimal interaction giving rise to 
second-order field equations without the Ostrogradski 
instability. The standard YM Lagrangian 
$-F^{a}_{\mu \nu}F^{a \mu \nu}/4$ supplemented by the above Horndeski 
non-minimal coupling gives rise to a Lagrangian whose simplicity makes it very appealing for cosmological applications. This model was already exploited 
in Ref.~\cite{Davydov} to develop an inflationary model. Similarly to gauge-flation, the Horndeski interaction allows for the existence of (quasi) de Sitter 
solutions with a graceful exit \cite{Davydov}, so this model is viable 
at the background level. While in gauge-flation the de Sitter expansion is supported by higher-order interactions, in the presence of the Horndeski interaction it is the non-minimal coupling which allows the inflationary solution. 

For the abelian case, the Horndeski interaction does not allow for inflationary solutions \cite{Barrow} and, in addition, ghosts and/or 
Laplacian instabilities typically arise in regions where the Horndeski non-minimal 
interaction dominates over the Maxwell term \cite{Jimenez13}. This poses a serious problem to any background cosmology relying on the Horndeski interaction for the abelian case. The existence of de Sitter solutions in the non-abelian case is already an improvement with respect to the abelian case. In this work, we will 
study whether the instability found for the abelian case can be avoided for 
the non-abelian case by exploring the stability of perturbations around 
the inflationary solution. Unfortunately, we will see that this is not the case and 
that instabilities persist in the non-abelian case.

The paper is organized as follows.
In Sec.\,\ref{nonabesec} we briefly review the background dynamics 
to show the existence of de Sitter solutions and we also confirm this numerically. 
In Sec.\,\ref{persec} we compute the quadratic action for tensor perturbations and show that they 
are plagued by ghosts and Laplacian 
instabilities on the inflationary background. We also show that ghost instabilities in the tensor sector
appear in more general cosmologies based on the Horndeski non-minimal interaction.
We then conclude  that our analysis of tensor perturbations alone is sufficient to 
exclude the HYM theory as a viable model of inflation and that strong coupling problems are expected in the vector and scalar sectors as well.

%%%%%%%%%%%%%%%%%%%%%%%%%%%%%%%%%%%%%%%%%%%%%%%%
\section{HYM theory and inflationary solutions}
\label{nonabesec}
%%%%%%%%%%%%%%%%%%%%%%%%%%%%%%%%%%%%%%%%%%%%%%%%

Let us begin with the HYM theory given by the action  
\be
S=\int d^4 x \sqrt{-\tilde{g}} \left( \frac{\mpl^2}{2}R -\frac{1}{4}F^{a\mu\nu}F^a_{\mu\nu} 
+\beta L^{\mu\nu\alpha\beta} F_{\mu\nu}^a F_{\alpha\beta}^a \right),
\label{action}
\ee
where $\tilde{g}$ is the determinant of the metric tensor $g_{\mu \nu}$, $M_{\rm pl}$ is 
the reduced Planck mass, $R$ is the Ricci scalar, $F^a_{\mu\nu}$ is the 
field strength tensor of a $SU(2)$ non-abelian gauge field $A^a_{\mu}$ 
(with $a=1,2,3$) defined by Eq.~(\ref{Fmunu}), $L^{\mu\nu\alpha\beta}$ 
is the double dual Riemann tensor given by Eq.~(\ref{double}), and 
$\beta$ is a coupling constant with dimension [mass]$^{-2}$.

As in the gauge-flation scenario \cite{gaugeinf},
the non-abelian gauge field configuration consistent with 
the flat homogenous and isotropic cosmological background
with the line element $ds^2=-dt^2 
+a^2(t) \delta_{ij}dx^i dx^j$ (where $a(t)$ is the scale factor) 
is given by 
\be
A_0^a= 0\,, \qquad
A_i^a= a(t) A(t) \delta_i^a\,.
\label{A-back}
\ee
The time-dependent function $A(t)$ characterizes the strength 
of the spatial components of the vector field. 
Then, the non-vanishing field strength components read
\be
F_{0i}^a= a\left(\dot{A}+HA \right) 
\delta_i^a\,,\qquad 
F_{ij}^a=-ga^2 A^2 \epsilon^a{}_{ij}\,,
\ee
where a dot represents a derivative with respect to $t$, and 
$H \equiv \dot{a}/a$ is the Hubble expansion rate.
The two independent background equations of motion 
following from the variations of (\ref{action}) with respect to 
the vector field and the metric tensor
are given by 
\ba
\hspace{-1.0cm}
& &
3\mpl^2H^2-\frac32\left[g^2A^4+(\dot{A}+HA)^2\right] \nonumber \\
\hspace{-1.0cm}
& &
+24 \beta H\left[2g^2A^3(2\dot{A}+HA)+3H(\dot{A}+HA)^2\right]
=0\,,\label{eq00}\\
\hspace{-1.0cm}
& &
\left[16\beta \left\{2g^2A^3+H(2\dot{A}+3HA)\right\}-A\right]\dot{H} 
\nonumber \\
\hspace{-1.0cm}
& &
+(16 \beta H^2-1)(\ddot{A}+3H\dot{A}+2H^2A+2g^2A^3)=0\,.
\label{eqA}
\ea

Solving Eq.~(\ref{eq00}) for $\dot{A}$, it follows that 
\be
\dot{A}=  \frac{HA(1-32g^2 \beta A^2-48\beta H^2) 
\pm \sqrt{\cal G}}
{48 \beta H^2-1}\,,
\label{soleqdotA}
\ee
with the short-cut notation
\ba
{\cal G} &\equiv& (48\beta H^2-1) \left[ g^2 A^4 (32 \beta H^2+1)
-2M_{\rm pl}^2 H^2\right]  \nonumber \\
& &+1024g^4 \beta^2 H^2 A^6\,.
\ea
We require that ${\cal G}>0$ for
the existence of real solutions of $\dot{A}$.

From Eq.~(\ref{eqA}) the de Sitter solution characterized by 
$H=H_{\rm dS}={\rm constant}$ needs to obey 
\be
\left(16 \beta H_{\rm dS}^2-1 \right)
\left( \ddot{A}+3H_{\rm dS}\dot{A}+2H_{\rm dS}^2A
+2g^2A^3 \right)=0\,. 
\label{eqA-dS}
\ee
Since the branch $\ddot{A}+3H_{\rm dS}\dot{A}+2H_{\rm dS}^2A+2g^2A^3=0$ gives rise to only the trivial solution $A=0$,\footnote{One can see this by differentiating Eq.~\eqref{eq00} with respect to time and taking the de-Sitter limit. This provides another equation that contains $\ddot{A}$. In general, this equation and the current branch of Eq.~\eqref{eqA-dS} are independent, and combined 
with Eq.~\eqref{soleqdotA}, one can solve algebraic equations to find $A$ as a function of $H_{\rm dS}$ alone, i.e., $A = {\rm constant}$. This leads to $2 H_{\rm dS}^2 A + 2 g^2 A^3 = 0$, which only has a trivial solution $A = 0$.} 
the de Sitter solution is uniquely fixed to be 
\be
H_{\rm dS}=\frac{1}{4\sqrt{\beta}}=\frac{1}{4}\mu\,,
\label{HdS}
\ee
where we introduced a mass scale $\mu$ related to 
$\beta$, as $\beta=1/\mu^2$. The constant $\beta$ 
must be positive for the existence of de Sitter solutions. 
From the observed amplitude of CMB temperature anisotropies we obtain an upper bound for the tensor-to-scalar ratio $r$. For a large class of models of inflation, this bound translates into an upper bound for the scale of inflation characterized by
$H_{\rm dS} \lesssim 10^{14}$~GeV \cite{Planckinf,TOKD}.
Thus, in the following we will consider the case in which $\mu$ 
is at most of the order of $10^{14}$~GeV.

There are two possibilities for the evolution of $A(t)$ along the 
de Sitter solution (\ref{HdS}). 
The first possibility is the constant $A$ solution, under which 
$\dot{A}=0$ in Eq.~(\ref{eq00}). This solution needs to satisfy 
the relation  
$A^2=(H_{\rm dS}^2/g^2)
\left[ -1 \pm \sqrt{1-2g^2 M_{\rm pl}^2/H_{\rm dS}^2} 
\right]$, but the r.h.s. becomes imaginary or negative.
In this case, there are no viable de Sitter solutions.

The second possibility is de Sitter solutions with time varying $A$. 
Since the r.h.s. of Eq.~(\ref{soleqdotA}) does not vanish, 
these solutions are not fixed points but temporal de Sitter attractors. 
Substituting Eq.~(\ref{HdS}) into Eq.~(\ref{soleqdotA}), we obtain
\be
\dot{A}=-\frac{g^2 A^3}{H_{\rm dS}} \left[ 1+
\epsilon \pm 
\sqrt{1+\frac{3}{2}\epsilon
-\epsilon^2 \left( 
\frac{M_{\rm pl}}{A} \right)^2} \right]\,,
\label{dotAds}
\ee
where 
\be
\epsilon \equiv \left( \frac{H_{\rm dS}}{gA} \right)^2\,.
\ee
Let us consider the regime $\epsilon \ll 1$, as the opposite case $\epsilon \gg 1$ with $A / M_{\rm pl} \lesssim \sqrt{\epsilon}$ would give no solution for $\dot{A}$ from Eq.~\eqref{dotAds} (equivalent to ${\cal G} < 0$). Then the plus branch of Eq.~\eqref{dotAds} approximately reduces to $\dot{A} \simeq -2(g^2/H_{\rm dS}) A^3$, integrated to give the solution $A(t) \simeq \sqrt{H_{\rm dS}/[4g^2(t-t_0)]}$, where $t_0$ is an integration constant. This solution is valid only for a short time period, $t-t_0 \ll H_{\rm dS}^{-1}$. As shown in Ref.~\cite{Davydov}, this branch of solutions is not an attractor, and even if one tunes initial conditions to momentarily realize this solution, the system quickly deviates away from it, with the time scale $\sim \epsilon H_{\rm dS}^{-1} \ll H_{\rm dS}^{-1}$. It is unable to sustain inflation for a sufficient duration, and therefore we do not consider this branch any further.

%%%%%%%%%%%%%%%%%%%%%%%%%%%%%%
\begin{figure}
\begin{center}
\includegraphics[height=3.0in,width=3.2in]{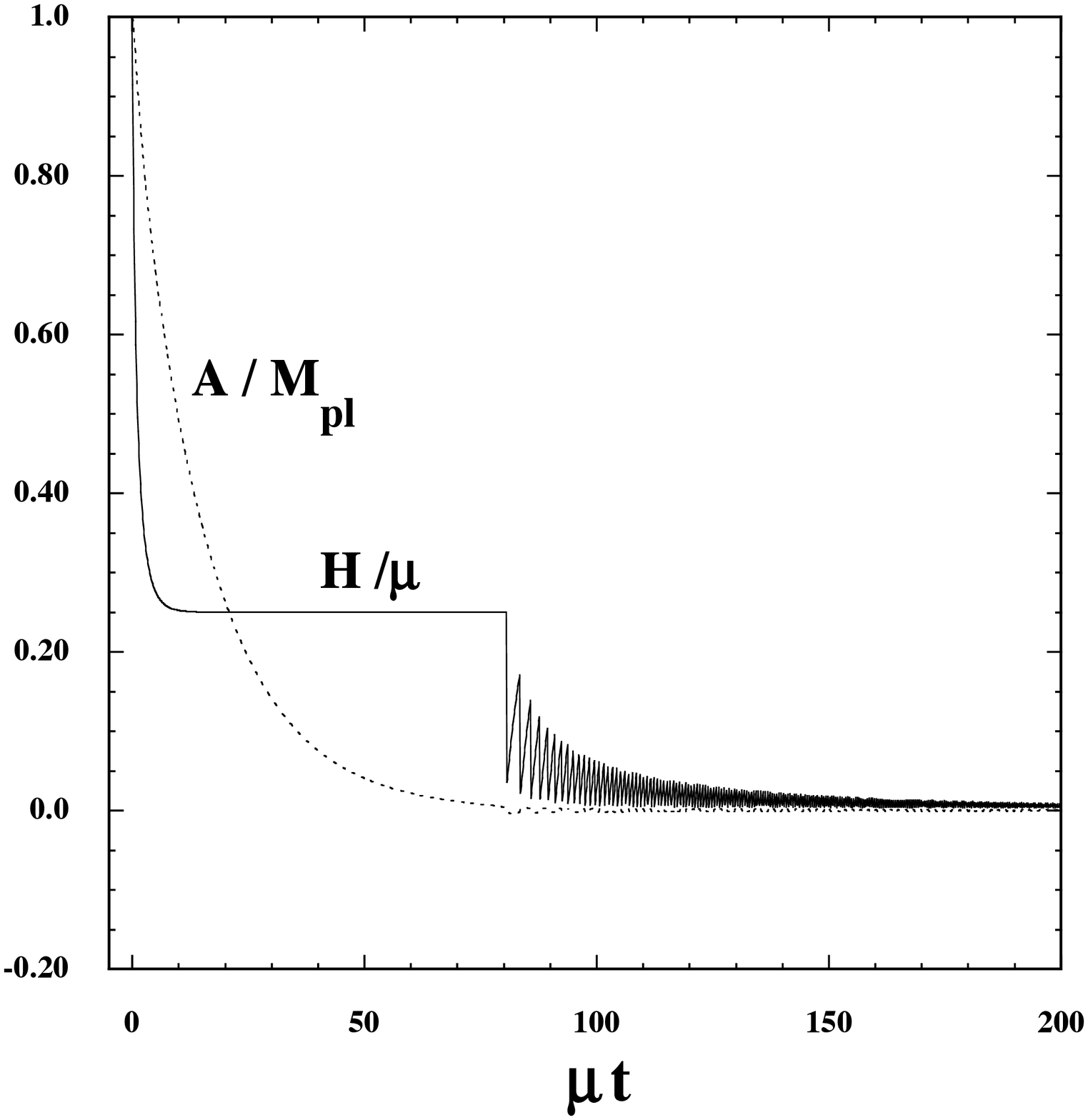}
\includegraphics[height=3.0in,width=3.3in]{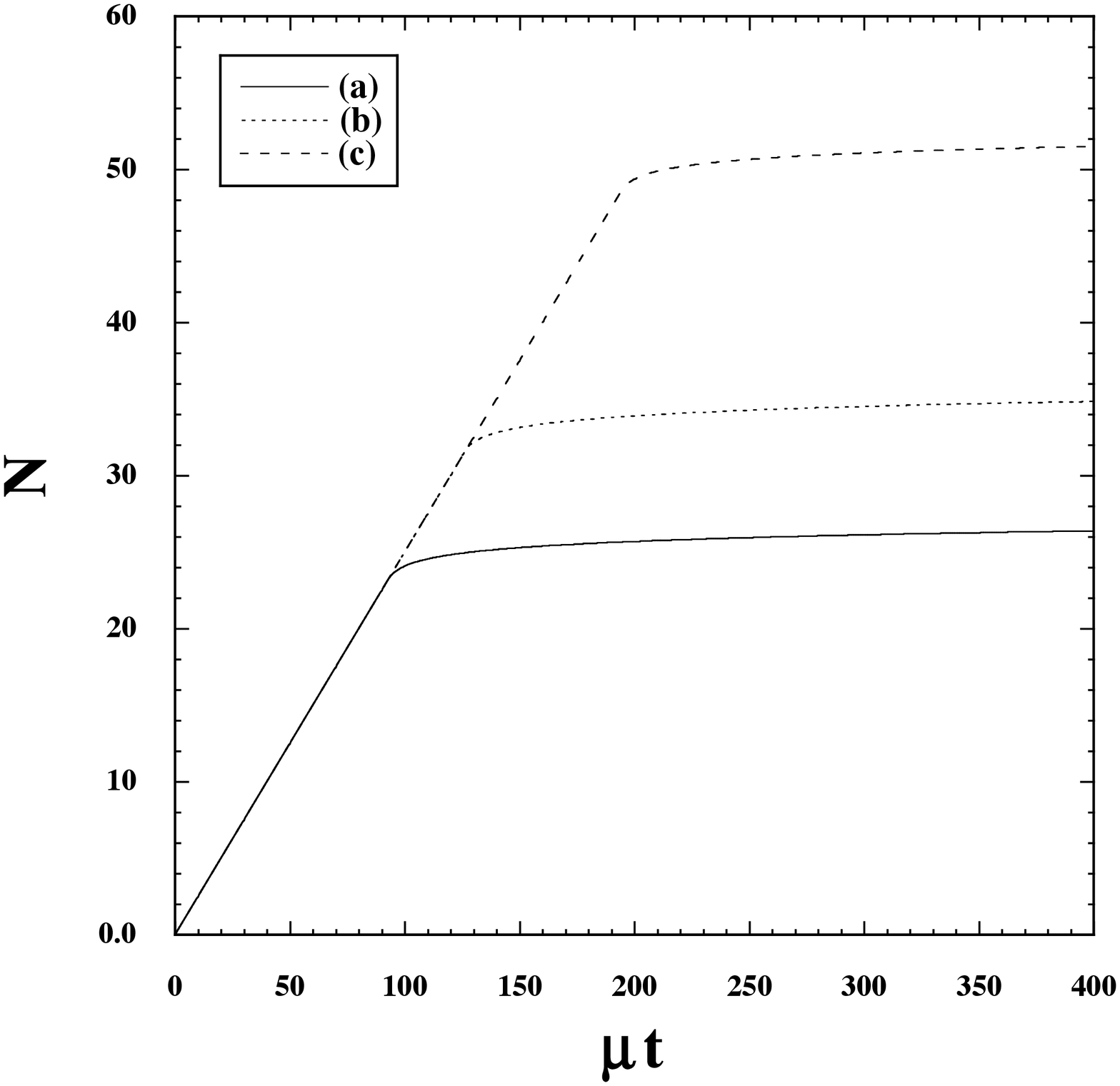}
\end{center}
\caption{\label{fig1}
(Top)
Evolution of $H/\mu$ and $A/M_{\rm pl}$  for 
$g=1$ and $\mu=10^{-4}M_{\rm pl}$. 
The initial conditions are chosen to be
$A_i=M_{\rm pl}$, $\dot{A}_i=0$, and $H_i=\mu$.
(Bottom) Evolution of the number of e-foldings 
in three different cases: 
(a) $\mu=10^{-4}M_{\rm pl}$ and the initial field 
value $A_i=M_{\rm pl}$, 
(b) $\mu=10^{-6}M_{\rm pl}$ and 
$A_i=M_{\rm pl}$, 
(c) $\mu=10^{-6}M_{\rm pl}$ and $A_i=10^2M_{\rm pl}$. 
Other model parameters and initial conditions are 
chosen to be $g=1$, $\dot{A}_i=0$, and 
$H_i=\mu/4$.}
\end{figure}
%%%%%%%%%%%%%%%%%%%%%%%%%%%%%%

Under the condition $\epsilon \ll 1$ the minus branch 
of Eq.~(\ref{dotAds}) reduces to 
$\dot{A} \simeq -H_{\rm dS}A/4$, 
so we obtain the integrated solution 
\be
A(t) \simeq A_0e^{-H_{\rm dS} t/4}\,,
\label{Aso}
\ee
where $A_0$ is a constant. 
As shown in Ref.~\cite{Davydov}, this is 
a temporally stable de Sitter solution with a graceful exit.
As $A(t)$ decreases during inflation, the last term 
in the square root of Eq.~(\ref{dotAds}) grows in time. 
The end of inflation is characterized by the epoch 
at which this term is comparable to 
the second term in the square root 
of Eq.~(\ref{dotAds}), so the gauge-field value 
at the end of inflation can be estimated as
\be
A_f \simeq 0.45 \sqrt{\frac{\mu M_{\rm pl}}{g}}\,.
\label{Af}
\ee
For concreteness, we consider the case in which $g$ is of order ${\cal O}(1)$. 
The initial value of $A$ must be larger than the one given in Eq.~\eqref{Af}, 
and $\epsilon \ll 1$ is satisfied during the whole stage of inflation for 
$H_{\rm dS} \lesssim 10^{-4} M_{\rm pl}$.

In the top panel of Fig.~\ref{fig1} we show the evolution of 
$H$ and $A$  for $g=1$ and $\mu=10^{-4}M_{\rm pl}$ 
with the initial conditions $A_i=M_{\rm pl}$, $\dot{A}_i=0$, 
and $H_i=\mu$. The Hubble parameter quickly approaches 
the value $H=\mu/4$, so the de Sitter solution (\ref{HdS}) 
is indeed a stable temporal attractor. During this de Sitter stage,
the gauge field $A$ decreases according to the analytic 
solution (\ref{Aso}). {}From Eq.~(\ref{Af}) the field 
value at the end of inflation can be estimated as 
$A_f=4.5 \times 10^{-3}M_{\rm pl}$.
In fact, our numerical simulation shows that, after $A$ 
drops below the order of $10^{-3}M_{\rm pl}$, 
the solution exits from the de Sitter period 
to enter the oscillating stage of the gauge field.

The number of e-foldings from the onset of inflation 
($t=t_i$) to the end of inflation is defined 
by $N=\int_{t_i}^{t_f}H dt$. 
On the de Sitter solution (\ref{HdS}) 
characterized by the field evolution (\ref{Aso}),
it follows that 
\be
N \simeq 
4 \ln \frac{A_i}{A_f}
\simeq 
4 \ln \frac{A_i}{\sqrt{\mu M_{\rm pl}}}
+2\ln g+3.2\,,
\label{Nestimate}
\ee
where $A_i$ is the initial value of $A$, and we used 
Eq.~(\ref{Af}) in the last approximate equality.

In the bottom panel of Fig.~\ref{fig1} we plot the evolution of 
$N$ for $g=1$ with several different values of $\mu$ 
and initial conditions of $A$.
The case (a) corresponds to $\mu=10^{-4}M_{\rm pl}$ 
and $A_i=M_{\rm pl}$, in which case the analytic 
estimation (\ref{Nestimate}) gives $N \simeq 22$. 
This shows fairly good agreement with the numerical 
value $N \simeq 23$. 
{}From Eq.~(\ref{Nestimate}) the number of 
e-foldings gets larger for smaller $\mu$ and larger 
$A_i$. In the case (b) of Fig.~\ref{fig1} the value of $\mu$ 
is $10^{-6}M_{\rm pl}$ with $A_i=M_{\rm pl}$, 
in which case $N \simeq 31$ from Eq.~(\ref{Nestimate}).
To realize inflation with $N>60$ (as required to solve 
flatness and horizon problems) for $A_i$ of the order of 
$M_{\rm pl}$, the Hubble scale $\mu$ needs to be 
smaller than $10^{-12}M_{\rm pl} \approx 10^6$~GeV 
(corresponding energy scale is $\sim \sqrt{\mu M_{\rm Pl}} \sim 10^{12} \, {\rm GeV}$). 
This is an unusual low-energy scale inflation. 

As far as the background evolution is concerned, the only relevant parameter, besides initial conditions, is the value of $g M_{\rm pl} / \mu$. Therefore, if one unfixes the value of $g$ and attempts to realize sufficiently long inflation with $A_i \sim M_{\rm pl}$ and $\mu \sim 10^{-4} M_{\rm pl}$, then $g \gg 1$ is necessary, 
which is also unfavorable in a perturbative point of view.
Another possibility for realizing the sufficient 
amount of e-foldings is to choose $A_i$ larger 
than the order of $M_{\rm pl}$. 
The case (c) of Fig.~\ref{fig1}, 
which corresponds to $\mu=10^{-6}M_{\rm pl}$ and 
$A_i=10^{2}M_{\rm pl}$, gives rise to the value 
$N \simeq 50$. This means that $A_i$ must be much 
larger than $M_{\rm pl}$ to realize $N>60$ for 
the standard energy scale of inflation 
($\mu \gtrsim 10^{-6}M_{\rm pl}$).

The above argument shows that either very low-energy 
scale inflation, a value of $g \gg 1$, or a very large initial 
value of $A_i$ is required for non-abelian gauge-flation 
to address the flatness and horizon problems. Although unusual, 
these features do not prevent the model from providing a successful 
inflationary scenario. 
A more serious flaw concerning the stability of perturbations 
will be faced in the next section.

%%%%%%%%%%%%%%%%%%%%%%%%%%%%%%
\section{Tensor perturbations}
\label{persec}
%%%%%%%%%%%%%%%%%%%%%%%%%%%%%%

In this section we will turn to the study of the perturbations around the de Sitter background discussed above. We will focus on the tensor sector where, 
in addition to the usual gravitational waves, the ``tensor'' part of the gauge fields is also 
present. More explicitly, the tensor perturbations in the metric tensor
are expressed in the form $\delta g_{ij}=a^2(t)h_{ij}$, 
where $h_{ij}$ satisfies transverse and traceless conditions 
$\partial^i h_{ij}=0$ and ${h^i}_i=0$. The perturbations of 
the non-abelian gauge-field $A_{\mu}^a$ can also be decomposed into 
``tensor'', ``vector'', and ``scalar'' modes \cite{gaugeinf}.
We write the gauge-field perturbations of the tensor part
as $\delta A_{i}^{a}=a(t)\gamma^a_{i}$ with the transverse and traceless conditions
$\delta^i{}_a\partial_i\gamma^a{}_j=0$ and $\delta^i{}_a{\gamma^a}_i=0$, 
so that $\gamma^a{}_{i}$ propagates the corresponding two polarizations. 
Notice that some of the contractions in the transverse and traceless conditions are done in terms of the $SU(2)$ indices. This implies that this decomposition does not have the standard interpretation with respect to spatial rotation, but it is consistent with the background configuration \eqref{A-back} and is useful in that each sector is decoupled from each other and coupled separately to the true scalar, vector and tensor sectors in the metric perturbations.

Choosing a coordinate system where the momentum ${\bm k}$ is 
oriented along the positive $z$-direction, the four tensor modes
mentioned above can be expressed in terms of the four 
functions $h_+(t,z)$, $h_{\times}(t,z)$, 
$\gamma_+(t,z)$, $\gamma_{\times}(t,z)$, as
\ba
\hspace{-1cm}
&&\delta g_{11}=
-\delta g_{22}=a^2h_+\,,\qquad \delta g_{12}
=\delta g_{21}=a^2h_{\times},\\
\hspace{-1cm}
&&\delta A^1_{\mu}=a(0,\gamma_+,\gamma_{\times},0)\,,\qquad 
\delta A^2_{\mu}=a(0,\gamma_{\times},-\gamma_{+},0)\,. 
\ea 
Now, we expand the action (\ref{action}) up to second order 
in tensor perturbations. 
After integrating the terms $\ddot{h}_+$ and $\ddot{h}_\times$ 
by parts, the second-order kinetic action containing the products of 
first-order time derivatives is given by 
\be
S_K^{(2)}=
\int \d^3x\d t\,a^3\,\dot{\vec{x}}^{\rm T}
{\bm K}\dot{\vec{x}}\,,
\label{SK}
\ee
where $\vec{x}^{\rm T}=(M_{\rm pl}h_{+},\gamma_{+},M_{\rm pl}h_{\times},\gamma_{\times})$, and ${\bm K}$ is the 4$\times$4 symmetric matrix 
whose non-vanishing 
components are given by 
\ba
&&K_{11}=\frac{1}{4}-\frac{2\beta}{M_{\rm pl}^2} 
\left[2g^2A^4-(\dot{A}+HA)^2\right]\,,\\
&&K_{12}=K_{21}=\frac{8\beta}{M_{\rm pl}} 
\left[g^2A^3+H(\dot{A}+HA)\right]\,,\\
&&K_{22}=1-16 \beta H^2\,,
\ea
and $K_{33}=K_{11}$, $K_{34}=K_{43}=K_{12}$, $K_{44}=K_{22}$.
Since the kinetic action (\ref{SK}) is the sum of two separate polarization states 
($+$ and $\times$), we should obtain two degenerate 
eigenvalues $\lambda_{\pm}$ for the matrix ${\bm K}$. 
The components $K_{22}$ and $K_{44}$ vanish 
on the de Sitter solution (\ref{HdS}), so it follows that 
\be
\lambda^{\rm dS}_{\pm}=\frac{1}{2} \left( K_{11} 
\pm \sqrt{K_{11}^2+4K_{12}^2} \right)\,.
\ee
From this expression we can easily see that the two eigenvalues will always have different signs (the product of the two gives $-K_{12}^2<0$) and, thus, the presence of a ghostly tensor sector along the de Sitter solution is unavoidable. In our case $K_{11}$ is positive, so that the eigenvalue $\lambda^{\rm dS}_{+}$ is positive and $\lambda^{\rm dS}_{-}$ is negative.

The presence of the ghost instability already renders the model unviable, but in the following we will see that the stability is further compromised due to the presence of Laplacian instabilities for modes well inside the Hubble horizon.
Besides the kinetic action (\ref{SK}), the dominant contributions 
to the total action in the small-scale limit ($k \gg aH$) 
are the terms containing the products of first-order 
spatial derivatives and the products of 
first-order time and spatial derivatives. 
After integration by parts, the second-order tensor action 
corresponding to those contributions can be schematically 
expressed as\footnote{The second term in this expression with only one spatial derivative must be interpreted as standing for the interactions coming from the Horndeski coupling as, e.g., 
$\sim \epsilon^{ijk}\dot{h}_{im}\partial_j\gamma_k{}^m$, 
where the $\epsilon$ is due to the structure constants. }
\be
S_{LC}^{(2)}=\int \d^3x\d t
\left( -a\,\partial \vec{x}^{\rm T}{\bm L} \partial {\vec x}
+a^2\,\dot{\vec{x}}^{\rm T}{\bm C} \partial {\vec x} 
\right)\,,
\label{SL}
\ee
where $\partial$ is derivative with respect to $z$, and 
${\bm L}$ and ${\bm C}$ are the $4 \times 4$ symmetric matrices 
with non-vanishing components
\ba
& &
L_{11}=
\frac{1}{4}+\frac{6\beta}{M_{\rm pl}^2} 
\left( \dot{A}+HA \right)^2\,,
\label{L11}\\
& &
L_{12}=L_{21}=
\frac{8\beta}{M_{\rm pl}} 
\left( \ddot{A}+H\dot{A}+\dot{H}A \right)\,,\\
& &
L_{22}= 1-16\beta \left(H^2+\dot{H} \right)\,,
\label{L22}
\ea
with $L_{33}=L_{11}$, $L_{34}=L_{43}=L_{12}$, 
$L_{44}=L_{22}$, and 
\be
C_{23}=C_{32}=-C_{14}=-C_{41}
=\frac{24\beta}{M_{\rm pl}} g A \dot{A}\,.
\ee
The dispersion relation can be derived by substituting a solution 
of the form $\vec{x}^{\rm T}=\vec{x}_0^{\rm T}e^{i(\omega t-kz)}$ 
into the equations of motion of tensor perturbations.
Defining the tensor propagation speed $c_t$ as $\omega=c_t k/a$, 
it needs to satisfy the relation ${\rm det} (c_t^2{\bm K}-{\bm L}-c_t{\bm C})=0$ 
for the existence of non-trivial solutions. 
On the de Sitter background (\ref{HdS}) the components 
$L_{22}$ and $L_{44}$ vanish, so $c_t^2$ obeys 
$[K_{12}^2c_t^4+(C_{23}^2-2K_{12}L_{12})c_t^2+L_{12}^2]^2=0$. Again, this equation will give rise to two degenerate solutions whose expressions are cumbersome and not very illuminating. Instead of displaying the full expressions, we will use the fact 
that the de Sitter solution exists for a sufficiently long period 
under the condition $\epsilon \ll 1$. 
In that limit, the squared propagation speeds reduce to
\be 
c_{t,\pm}^2 \simeq \frac{-15 \pm 3\sqrt{21}}{32}\epsilon,
\ee
which are both negative. Hence, as advertised, the tensor modes 
not only contain ghost instabilities, but they are also prone to 
Laplacian instabilities.

We will end this section by showing when the tensor ghost arises
without assuming the de Sitter solution discussed above.
In doing so, we resort to the fact that the product of two different eigenvalues 
of the kinetic matrix ${\bm K}$ is given by 
$\lambda \equiv \lambda_{+}\lambda_{-}=K_{11}K_{22}-K_{12}^2$.
On using the background equation (\ref{soleqdotA}) to eliminate 
the derivative $\dot{A}$ in $K_{11}$ and $K_{12}, K_{21}$, we obtain
\be
\lambda
=\frac{1}{4}-2X^2 \xi \left( 32\xi +3\right)\,,
\label{lampm}
\ee
where we have introduced $\xi \equiv \beta g^2 A^2$ and 
$X \equiv A/\mpl$.
For the absence of ghosts the condition $\lambda>0$ is necessary, 
but it is not sufficient. However, having $\lambda<0$ 
necessarily implies the existence of ghosts.

On the de Sitter solution we have $\xi=1/(16\epsilon) \gg 1$ and 
$X \gtrsim 1$, so the expression (\ref{lampm}) reduces to 
$\lambda \simeq -64 \xi^2 X^2<0$. 
In the opposite regime satisfying the conditions $\xi \ll 1$ and $X \lesssim 1$ 
the quantity $\lambda$ is close to $1/4$, 
so it is possible to avoid the appearance of tensor ghosts. 
Since the constant factor $1/4$ in Eq.~(\ref{lampm}) arises from the kinetic term 
of the YM Lagrangian $-F^{a \mu \nu}F^a_{\mu \nu}/4$, the disappearance of ghosts in this case corresponds to the regime where the YM kinetic term dominates over the Horndeski non-minimal coupling $\beta L^{\mu\nu\alpha\beta} F_{\mu\nu}^a F_{\alpha\beta}^a$. 

If the Horndeski interaction dominates over the YM term, we expect to have $\lambda\simeq-2X^2\xi(32\xi+3)$ which can only be positive if $-3/32<\xi<0$. This definitely rules out stable cosmologies driven by the Horndeski coupling that are continuously connected with de Sitter solutions, which only exist for $\xi>0$. We should stress again that the condition $\lambda>0$ is not sufficient to avoid ghosts, but we have checked that the values of $\xi$ in the range $-3/32<\xi<0$ can lead to ghost-free cosmologies (although some other pathologies might be present). In any case, the very narrow range obtained for $\xi$ clearly shows that, quite generically, the Horndeski coupling will give rise to ghost instabilities whenever it dominates. This result is analogous to what was found for the abelian 
case in Ref.~\cite{Jimenez13}. 

For $\beta \to 0$ the components of ${\bm K}$ 
reduce to $K_{11} \to 1/4$, $K_{12}=K_{21} \to 0$, and $K_{22} \to 1$, so there are no ghosts in this limit, as a consistency check that a free non-abelian field exhibits no pathological 
behaviors. We also note that, as $A$ and $H$ decrease towards 0 
during reheating, the solutions enter the regime 
in which the tensor ghosts are absent. 
This corresponds to the radiation-dominant epoch
driven by the YM term (which dominates over the 
Horndeski coupling).

%%%%%%%%%%%%%%%%%%%%%%%%%%%%%%%%%%%%%%
%%%%%%%%%%%%%%%%%%%%%%%%%%%%%%%%%%%%%%%

%%%%%%%%%%%%%%%%%%%%%
\section{Conclusions}
%%%%%%%%%%%%%%%%%%%%%

In this work, we have considered the model of inflation based on a non-abelian gauge field with the Horndeski non-minimal coupling. We have briefly reviewed the existence of inflationary background solutions and studied the property of  tensor perturbations.
We have found that these tensor modes are plagued by ghosts and small-scales Laplacian instabilities on de Sitter solutions. The presence of these instabilities in the model casts 
serious problems on the viability of the HYM inflationary scenario, since the background will quickly be spoilt by small scale perturbations and the usual quantization procedure of tensor perturbations as the initial Bunch-Davies vacuum state will no longer be valid \cite{MFB}. Finally, we have found that, 
beyond de Sitter solutions, the presence of ghosts in the tensor sector persists
whenever the Horndeski non-minimal coupling dominates over the YM term.

The reason why the ghost instabilities arise for the tensor modes roots in the own inflationary solution supported by the non-minimal coupling. In fact, for the de Sitter background, the double dual Riemann tensor can be expressed as $L^{\mu\nu\alpha\beta}=2H_{\rm dS}^2\big(g^{\mu\alpha}g^{\nu\beta}-g^{\nu\alpha}g^{\mu\beta}  \big)$ so that the HYM Lagrangian density becomes 
\be
{\mathcal L}_{\rm HYM}^{\rm dS}=-\frac14\Big(1-16\beta H_{\rm dS}^2\Big)F_{\mu\nu}^a F^{a\mu\nu}\,,
\ee
which exactly vanishes along the de Sitter solution and, thus, we can expect some problems arising from here. For the tensor modes, we can see that the Einstein-Hilbert term will contribute the usual $\dot{h}^2$ to the time derivatives of the metric perturbations. On the other hand, the HYM Lagrangian will only contribute terms of the forms $\dot{h}^2$ and $\dot{h}\dot{\gamma}$, while the terms with $\dot{\gamma}^2$ will drop for the de Sitter solution. Thus, the kinetic matrix will acquire the following simple form 
\[{\bm K}\sim
\left(
\begin{array}{cc}
 a &b      \\
 b &0      \\ 
\end{array}
\right)
\]
for some time-dependent quantities $a$ and $b$. 
The eigenvalues of this matrix are $\lambda_{\pm}=(a\pm\sqrt{a^2+4b^2})/2$ and, thus, one mode will always be a ghost, as we obtained in a more rigorous way above. Moreover, this analysis also allows us to identify the presence of strong coupling issues for the vector and scalar sectors on the de Sitter solution. For those sectors, all the kinetic terms must arise from the HYM Lagrangian since the Einstein-Hilbert action only contributes kinetic terms for the tensor modes. Thus, the kinetic terms for the vector and scalar modes will be preceded by the factor $1-16\beta H_{\rm dS}^2$ that vanishes along the de Sitter solution and, thus, one would expect that the strong coupling problem arises. 
While one might worry that the actual situation is not as simplistic as this argument due to the presence of non-dynamical modes in the vector and scalar sectors that should be integrated out before instability analyses, we
have checked this explicitly by computing the corresponding quadratic 
actions and have indeed found these strong couplings. 

We conclude that, despite being very appealing due to its simplicity, the HYM Lagrangian seems to share the pathologies of its abelian counterpart, i.e., the appearance of instabilities whenever the Horndeski non-minimal coupling dominates over the YM term. In particular, we have seen that the promising inflationary model proposed in Ref.~\cite{Davydov} unfortunately suffers from severe instabilities and strong coupling problems which render it unviable. 

In order to have a viable inflationary model based on non-abelian gauge fields, we are forced to include higher-order interaction terms for the YM field similar to those in gauge-flation \cite{gaugeinf}, since we know that  they allow for viable regions in the parameters space without ghosts and/or instabilities. Although theoretically viable, the simplest higher-order interaction terms are in tension with the CMB constraints on $n_s$ and $r$ \cite{Namba}. A slight extension of the model including an additional scalar field can alleviate this tension \cite{Dimastrogiovanni:2016fuu}.
It may be interesting to study whether the introduction of 
the Horndeski non-minimal coupling to the simplest realizations of gauge-flation 
could allow for the existence of parameter space consistent 
with both theoretical and observational constraints. We should point out, however, that the Horndeski coupling could only contribute a small correction to the dynamics (otherwise one would run into the stability problems discussed throughout this work).
 
Let us finally comment on other possibilities to have this class of inflationary scenarios. The most obvious extension of gauge-flation would be to consider a general function of the Lorentz- and $SU(2)$-scalars that one can construct. Another route is to abandon the gauge character of the models and consider more general interactions like those considered in Refs.~\cite{Allys:2016kbq,Jimenez:2016upj,Emami}.

%%%%%%%%%%%%%%%%%%%%%%%%%%%
\acknowledgments 
JBJ  acknowledges  the  financial support of A*MIDEX project 
(n ANR-11-IDEX-0001-02) funded by
the  ``Investissements d'Avenir" French Government program, 
managed by the French National Research Agency
(ANR),  MINECO  (Spain) project
FIS2014-52837-P and Consolider-Ingenio MULTIDARK
CSD2009-00064.
LH thanks financial support from Dr.~Max R\"ossler, 
the Walter Haefner Foundation and the ETH Zurich
Foundation. RK is supported by the Grant-in-Aid for 
Research Activity Start-up of the JSPS No.\,15H06635. 
RN is supported by the Natural Sciences and Engineering
Research Council (NSERC) of Canada and by the Lorne 
rottier Chair in Astrophysics and Cosmology at McGill.
ST is supported by the Grant-in-Aid for Scientific Research 
Fund of the JSPS Nos.~24540286, 
16K05359, and MEXT KAKENHI Grant-in-Aid for 
Scientific Research on Innovative Areas 
``Cosmic Acceleration'' 
(No.\,15H05890).
%%%%%%%%%%%%%%%%%%%%%%%%%%%%

%%%%%%%%%%%%%%%%%

\end{document}